\documentstyle[sprocl,amsmath,amssymb,epsfig]{article}

\def\be{\begin{equation}}
\def\ee{\end{equation}}
\def\bea{\begin{eqnarray}}
\def\eea{\end{eqnarray}}

\begin{document}

\begin{flushright}
{\bf  CERN-TH/99-07 
\\ UTHEP-98-1101
\\ January 1999}
\end{flushright}

\vspace{2mm}

\title{PRECISION CALCULATION OF \\
       BHABHA SCATTERING AT LEP$^\dag$}

\author{
  W. P\L{}ACZEK$^{ab}$,
  S. JADACH$^{bc}$,
  M. MELLES$^{d}$,
  B.F.L. WARD$^{efb}$,
  S.A. YOST$^{e}$ 
}

\address{
$^a$Institute of Computer Science, Jagellonian University, \\
     ul. Nawojki 11, 30-072 Cracow, Poland, \\
$^b$CERN, CH-1211 Geneva 23, Switzerland,\\
$^c$Institute of Nuclear Physics,
    ul. Kawiory 26a, 30-055 Cracow, Poland,\\
$^d$Department of Theoretical Physics, University of Durham,\\ 
    South Road, Durham City 3LE, England,\\
$^e$Department of Physics and Astronomy, The University of Tennessee, \\
    Knoxville, TN 37996-1200, USA,\\
$^f$SLAC, Stanford University, Stanford, CA 94309, USA\\
}

\vspace{2mm}

\address{
 $^\dag$Talk given by W. P\l{}aczek at IVth International Symposium \\
 on Radiative Corrections RADCOR 98, \\
 Barcelona (Catalonia, Spain), Sept.\ 8-12, 1998.
}

%
%
%
%
%
%
%
%
%


\maketitle\abstracts{
For the small-angle Bhabha-scattering process,
we consider the error budget for the calculation of the
LEP/SLC luminosity in the Monte Carlo event generator BHLUMI~4.04,
from the standpoint of new calculations of exact results for the
respective ${\cal O}(\alpha^2)$ photonic corrections in the context of
the Yennie-Frautchi-Suura exponentiation. We find that 
an over-all precision tag for the currently available program BHLUMI~4.04 
can be reduced from $0.11\%$ to $0.061\%$ at LEP1 and from $0.25\%$
to $0.122\%$ at LEP2.
For the large-angle Bhabha process, we present the Monte Carlo
program BHWIDE and compare its predictions with predictions
of other Monte Carlo programs as well as semi-analytical calculations.
}

\section{Introduction}

At LEP, for practical purposes, the process of Bhabha scattering, 
$e^+e^- \rightarrow e^+e^-$,
is divided into two classes
depending on the kinematical regions%
      \footnote{In the following we shall always consider the case 
                with both the $e^+$ and $e^-$ detected.}%
: the small-angle Bhabha (SABH) scattering, for the scattering angle 
$1^{\circ} \lesssim \theta_e \lesssim 6^{\circ}$,
and the large-angle Bhabha (LABH) scattering, 
for $\theta_e \gtrsim 10^{\circ}$.
In general, this process is mediated by both the $\gamma$ and $Z$ 
bosons, exchanged in both $s$- and $t$-channels. 
Thus, at the lowest order, there are 4 pure $s$-
and $t$-channel contributions and 6 interference terms 
(between $\gamma$ and $Z$ and between the $s$- and $t$-channels). 
At low angles, however, the Bhabha scattering is almost completely dominated 
($\gtrsim 99\%$ of a cross section) 
by the pure QED process of the $t$-channel $\gamma$-exchange
for which a very high accuracy in theoretical predictions can be achieved.
For this reason SABH was chosen at LEP1 and LEP2 for the luminosity
monitoring. 
At large angles, physical features of the Bhabha process at LEP1 and
LEP2 are very different, as different Feynman-diagram contributions  
dominate at these two energy regimes. At LEP1 energies, a dominating
role is played by the $s$-channel $Z$-exchange, so this process is used,
in parallel to other fermion-pair production, to determine
properties of the $Z$-boson as well as to measure 
other important electroweak (EW) parameters~\cite{lepybk96}.
At LEP2, LABH is dominated by the $t$-channel $\gamma$ exchange, so
at these energies it resembles more SABH than 
the other two-fermion processes. 
Therefore, LABH is not very useful at LEP2 for testing the EW
sector of the Standard Model (SM). It is considered mainly 
as a background for those processes that are of the main experimental 
interest at LEP2.
The $e^+e^-$-channel is investigated particularly in searches for
possible ``new physics'', such as SUSY, contact interactions, etc.

Our discussion is organized as follows. 
In Section~2, we concentrate on the SABH scattering and 
discuss the new error budget for the Monte Carlo (MC) program 
BHLUMI~4.04~\cite{bhl4:1996} based on the exact 
${\cal O}(\alpha^2)$ calculations.
The LABH process is discussed in Section~3, where we give a
brief description of the MC event generator BHWIDE~\cite{bhw:1997}  
and discuss its cross-checks and comparisons with those of other programs.
Finally, Section~4 contains our conclusions and outlook.

\section{Small-Angle Bhabha Scattering}

Currently, new luminometers at LEP~\cite{lumin:1996} have given
results on the luminosity process $e^+ e^- \rightarrow e^+ e^- +
n(\gamma)$ at experimental precision tags below $0.1\%$.  This
should be compared with the prediction by the Krak\'ow-Knoxville 
Collaboration in the Monte Carlo program 
BHLUMI~4.04~\cite{bhl4:1996}, wherein the
theoretical precision tag of $0.11\%$ is realized for this process
in the ALEPH SICAL-type~\cite{sical} acceptance. If one combines the
experimental results, one arrives at an experimental precision
of $\lesssim 0.05\%$. 
Evidently, for the final EW precision tests
data analysis for LEP1, it would be desirable to reduce the
theoretical precision tag on the luminosity cross section prediction,
at least to the comparable $0.05\%$-regime, in order 
not to obscure unnecessarily
the comparison between experiment and the respective Standard Model
of the electroweak interaction. With this as our primary motivation,
we examined the error budget arrived at in 
Refs.~\cite{lepybk96,arbuzov:1996} in view of recent exact results
impacting both the technical and physical precision of the errors
quoted in that budget.
More precisely, if one looks into the error budget shown in Table~1 of
Ref.~\cite{arbuzov:1996}, one sees that the largest contribution
is associated with the ${\cal O}(\alpha^2)$ photonic corrections,
which contribute $0.1\%$ in quadrature to the total $0.11\%$ quoted
for the total precision of the BHLUMI~4.04 prediction in these
references for the ALEPH SICAL-type acceptance. Accordingly, we have 
used the exact results of Refs.~\cite{1r1v:1996,2brem:1992,2brem:1993} and the
exact result of Ref.~\cite{frits:1988} to make a more realistic
estimate of the true size of this dominant error quoted in
Refs.~\cite{lepybk96,arbuzov:1996}.

In re-examining the photonic corrections used in BHLUMI 4.04
at the ${\cal O}(\alpha^2)$, which is the relevant order of the
corrections, one needs to look at the approximations made in the
matrix element used in the calculation encoded in the program
in comparison to available exact results. This will allow us
to re-assess the physical precision of the corresponding
part of the BHLUMI~4.04 matrix element, which is the exact
${\cal O}(\alpha^2)$ LL (leading-log) Yennie-Frautschi-Suura (YFS)
exponentiated matrix element. 
The phase-space integration of two hard photon emission in BHLUMI is exact
(the LL approximations are only in the matrix element).
Nevertheless, this four-body phase-space integration should be cross-checked
with another, independent, exact integration method.
This check, which we have recently completed, will 
allow us to give a more realistic estimate of the
technical precision of the realization of the corresponding aspect of the
matrix element in BHLUMI~4.04. The net result is a new estimate
of the total precision of the prediction of the luminosity cross section
by BHLUMI~4.04 at LEP1 and LEP2 energies. 

\begin{figure}
\center
\setlength{\unitlength}{0.1mm}
\begin{picture}(1600,500)
\put(-20,0){\makebox(0,0)[lb]{
\epsfig{file=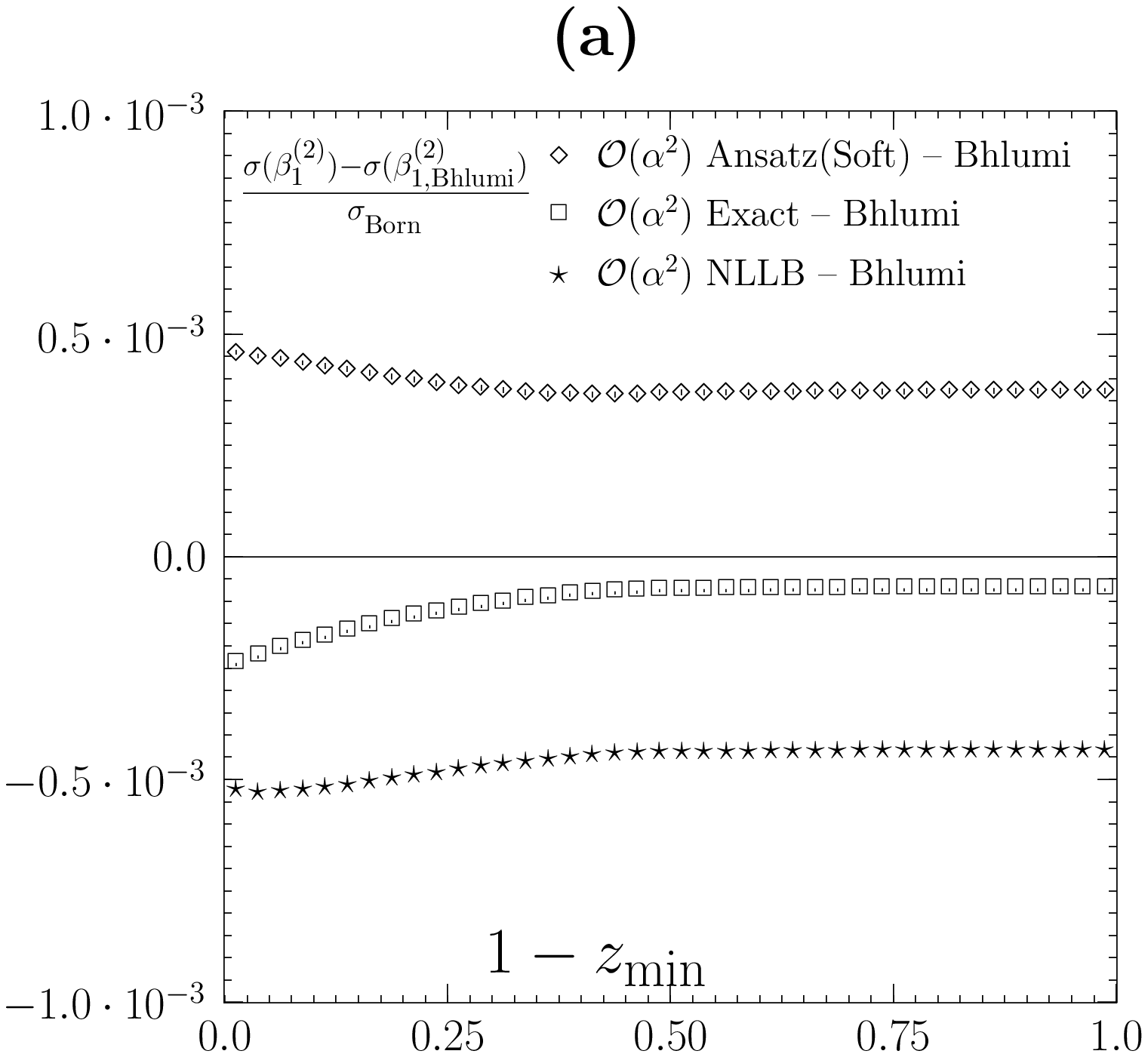,height=48mm,width=58mm}
}}
\put(600,0){\makebox(0,0)[lb]{
\epsfig{file=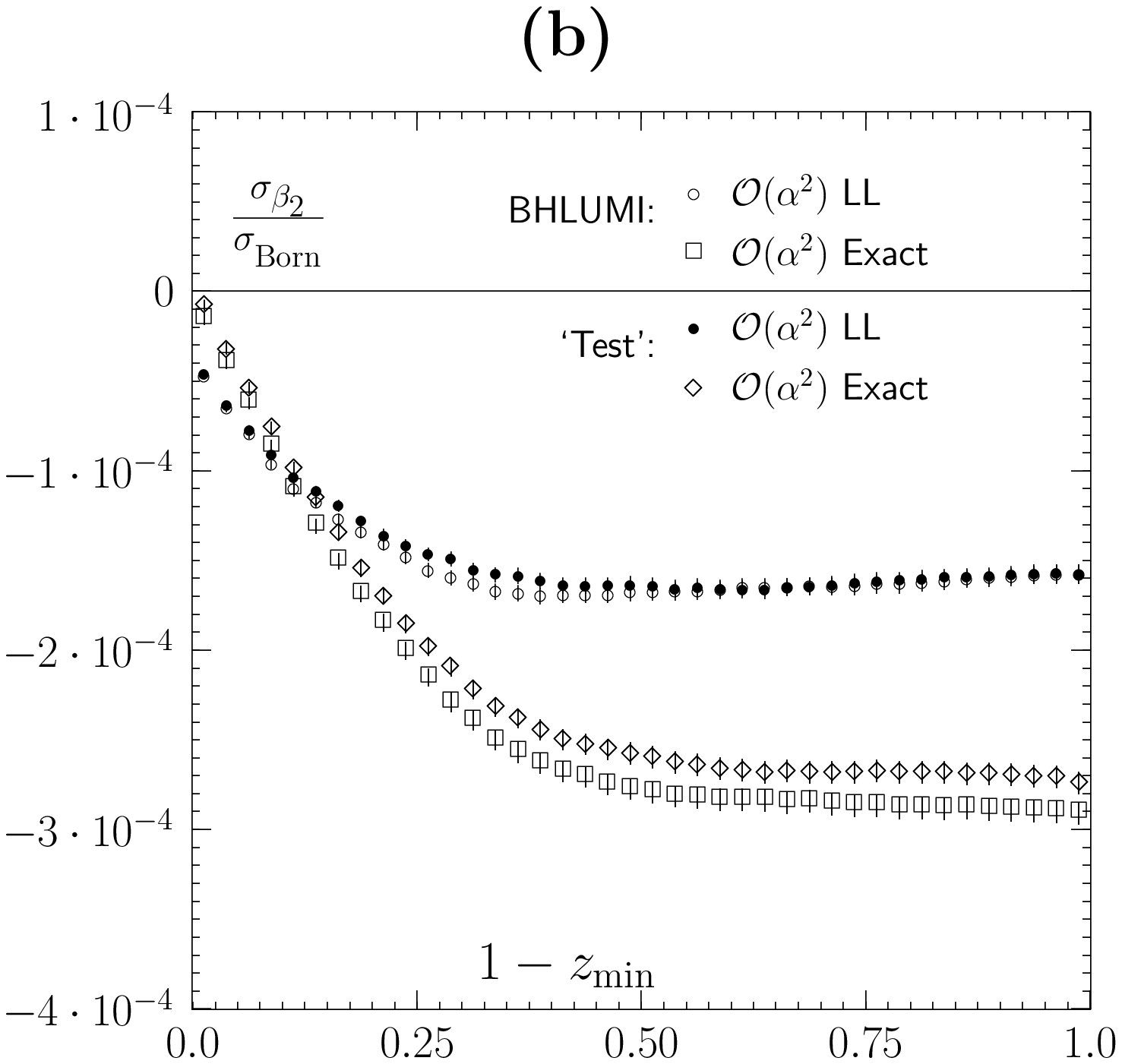,height=48mm,width=58mm}
}}
\end{picture}
\vspace{-7mm}
\caption{\small\sf
{\bf (a)} 
The pure second order Monte Carlo result for 
$\bar\beta^{(2)}_1- \bar\beta^{(2)}_{1,Bhlumi}$ 
differences for the SICAL Wide-Narrow
trigger, divided by the Narrow-Narrow Born cross section;
$z_{min}$ is as it is defined in Fig.~2 of Phys. Lett. {\bf B353} (1995) 362.
{\bf (b)}
Comparison of Monte Carlo results for $\bar\beta^{(2)}_2$
for the LL and exact matrix
elements. The results are shown for the BHLUMI generator
and for an alternative `Test' generator for a technical precision test.
}%
\label{fig:sabh}
\end{figure}

Considering now the exact ${\cal O}(\alpha)$ correction
to the single hard brems\-strah\-lung in the luminosity process,
we have implemented the results of Ref.~\cite{1r1v:1996} into BHLUMI~4.xx
and made a systematic study of the net change in the prediction
for the luminosity relative to the prediction of BHLUMI~4.04
in which this correction is treated to the LL level. What we find is
illustrated in Fig.~\ref{fig:sabh}a for the ALEPH SICAL-type acceptance
at the $Z$ peak. In the language of the YFS theory, this correction
enters the hard-photon residuals as $\bar\beta^{(2)}_1$, the 
${\cal O}(\alpha^2)$ contribution to the one-hard-photon residual
$\bar\beta_1$. 
In Fig.~\ref{fig:sabh}a, we show the
difference between the corresponding LL result in BHLUMI~4.04
and: (1) our exact result as given in Ref.~\cite{1r1v:1996},
(2) the approximate ansatz in Eq.~(3.25) of Ref.~\cite{1r1v:1996},
(3) the result (NLLB) of Ref.~\cite{EAK:1996}, which is supposed to
include the dominant non-LL effects, 
in ratio to the respective Born cross section.
What we see is that the BHLUMI~4.04 results are within $0.02\%$ 
of the exact result 
throughout the experimentally interesting regime $0.2\le 1-z_{min}\le 1.0$.
This is the main reason why we will be able to reduce the estimated precision
of the BHLUMI~4.04 prediction in comparison to 
Refs.~\cite{lepybk96,arbuzov:1996}.

Turning next to the technical precision of the $2\gamma$ 
bremsstrahlung calculation in BHLUMI~4.04, we have constructed a completely
independent realization of the two-photon phase-space integration 
compared to what is used in BHLUMI~4.04 by way of an independent 
MC algorithm. We have implemented this new MC
realization of the exact two-photon phase-space and compared its result
with that of BHLUMI~4.04's for the hard-photon residual $\bar\beta_2$
contribution to the luminosity cross section, both for the LL
matrix element in BHLUMI~4.04 and for the exact matrix element
for the two-photon bremsstrahlung of Refs.~\cite{2brem:1992,2brem:1993}. What
we find is shown in Fig.~\ref{fig:sabh}b for the ALEPH
SICAL-type acceptance at the $Z$-peak.
We find that the
difference between the two realizations of the exact $2\gamma$ bremsstrahlung
is below $0.003\%$ of the Born cross section. Moreover, we get
an estimate of the physical precision of the LL approximation for
this part of the cross section from comparing the LL and exact results
as $0.012\%$, in agreement with our estimate 
in Refs.~\cite{lepybk96,arbuzov:1996}.

Finally, we turn to the exact result for the two-loop contribution
of the hard-photon residual $\bar\beta_0$ to the cross section in comparison
to the LL result used for it in BHLUMI~4.04.  We have analytically continued
the result of Ref.~\cite{frits:1988} from the $s$-channel to the
$t$-channel for the required two-loop contribution to the respective
charge form factor in QED. In this way, using the YFS theory, we have
found that the difference between the LL result in BHLUMI~4.04
and the exact result corresponds to the shift of the function
$\upsilon$ in Eq.~(2) of Ref.~\cite{acta:1996} by 
{\small
\begin{center}
$\Delta\upsilon^{(2)} = \left(\frac{\alpha}{\pi}\right)^2 \left[ 
                         \left(6+6\zeta(3)-\frac{45}{8}
                        - \frac{\pi^2}{2}\right) L          
                      +  6-9\zeta(3)
                      +  \left(\frac{17}{8}-2\ln 2\right)\pi^2
                        - \frac{8}{45}\pi^4\right],         
$ \\
\end{center}%
}%
\noindent 
where the big logarithm is defined as $L=\ln |t|/m_e^2$ and $\zeta(3)$
is the Riemann $\zeta$-function of the argument $3$. For the ALEPH SICAL-type
acceptance at the $Z$-peak, this corresponds to $0.014\%$ in the 
cross section.

Collecting the above results in quadrature, we obtain the result
that the current calculation of the ${\cal O}(\alpha^2)$ photonic 
corrections in BHLUMI~4.04 are accurate to  ${\bf 0.027\%}$.
Using this result in Table~1 of Ref.~\cite{arbuzov:1996} we arrive at
the precision tag ${\bf 0.061\%}$ for the currently available calculation
in BHLUMI~4.04 at the $Z$ peak. At the LEP2 energy of $176$~GeV, if we
repeat the analysis just described, we find that the corresponding
precision of BHLUMI~4.04, for both the SICAL- and LCAL-type acceptances,
is now reduced to ${\bf 0.122\%}$, to be compared to the estimate 
of $0.25\%$ in Refs.~\cite{lepybk96,arbuzov:1996}. 
The current situation is now summarized in  Table~\ref{tab:total-error-lep}.
{\small
\begin{table*}[hbtp]
\centering
\begin{tabular}{|l|l|l|l|l|}
\hline
 & \multicolumn{2}{|c|}{LEP1} & \multicolumn{2}{|c|}{LEP2} \\
\hline
  Type of correction/error
& Past~\protect\cite{lepybk96,arbuzov:1996}
& Present
& Past~\protect\cite{lepybk96,arbuzov:1996}
& Present     \\
\hline
(a) Missing photonic 
    ${\cal O}(\alpha^2 )$~\protect\cite{elsewh} &
    0.10\%      & 0.027\%    & 0.20\%  & 0.04\%
\\
(b) Missing photonic 
    ${\cal O}(\alpha^3 L^3)$~\protect\cite{th-96-156} &
    0.015\%     & 0.015\%    & 0.03\%  & 0.03\%
\\
(c) Vacuum polarization~\protect\cite{burkhardt-pietrzyk:1995,eidelman-jegerlehner:1995} &
    0.04\%      & 0.04\%    & 0.10\%  & 0.10\%
\\
(d) Light pairs~\protect\cite{pairs:1993} &
    0.03\%      & 0.03\%    & 0.05\%  & 0.05\%
\\
(e) Z-exchange~\protect\cite{th-95-74}   &
    0.015\%      & 0.015\%   &  0.0\%  & 0.0\%
\\
\hline
    Total  &
    0.11\%      & 0.061\%    & 0.25\%  & 0.122\%
\\
\hline
\end{tabular}
\vspace{-2mm}
\caption{\sf
Summary of the total (physical+technical) theoretical uncertainty
for a typical
calorimetric detector.
For LEP1, the above estimate is valid for the angular range
within   $1^{\circ}$--$3^{\circ}$, and
for  LEP2  it covers energies up to 176~GeV, and the
angular ranges within $1^{\circ}$--$3^{\circ}$ and $3^{\circ}$--$6^{\circ}$.
}
\label{tab:total-error-lep}
\end{table*}
}
A more detailed exposition of the results in this paper will appear
elsewhere~\cite{elsewh}.

Our result on the size of the error associated with the 
missing sub-leading bremsstrahlung
correction at ${\cal O}(\alpha^2)$ in BHLUMI~4.04, which is $0.027\%$, agrees
with the estimate of $0.03\%$ made by Montagna {\it et al.}~\cite{oreste}, 
using a structure function convolution of a hard collinear external
photon with an acollinear internal photon. As these authors have argued,
while such a pairing of convolutions does not represent a complete set
of photonic ${\cal O}(\alpha^2L)$ corrections, one expects it to
contain the bulk of such corrections. Indeed, our exact result of
$0.027\%$ shows that the approximation made in Ref.~\cite{oreste}
does give the bulk of the respective ${\cal O}(\alpha^2L)$ correction.
Evidently, the fact that we now have two independent results, one exact, that
presented by us in this paper,
and one approximate, that in Ref.~\cite{oreste}, which agree on the
size of the error associated with the missing photonic 
${\cal O}(\alpha^2L)$ correction in BHLUMI~4.04, enhances the 
results in this paper.

\section{Large-Angle Bhabha Scattering}

At LEP1, the main physical quantities measured in LABH are:
the total cross section $\sigma_e$ and the forward--backward charge 
asymmetry ${\cal A}_{FB}$.
A value of $\sigma_e$ is used to extract the partial $Z$ decay
width $\Gamma_e$, while ${\cal A}_{FB}$ is sensitive to the important
EW parameters, such as the top and Higgs masses. 
The experimental precision for LABH, after the final LEP1 data analysis is
completed, is expected to be $\lesssim 0.5\%$ at the $Z$ peak and $\sim 1\%$ 
at $\pm 2$~GeV off the peak~\cite{benpas:1998}.
At LEP2, the $e^+e^-$-channel is considered mainly for the ``new physics''
searches, and there, the experimental precision is expected to reach 
$~0.5\%$.~\cite{tomalin}.
On the theory side, several programs (both the MC and semi-analytical 
ones) for LABH have been developed, see e.g. Ref.~\cite{lepybk96},
but a comprehensive analysis of the theoretical error is still missing%
\footnote{Recently, the analysis of the theoretical errors of two
          semi-analytical programs, ALIBABA and TOPAZ0, has been 
          presented~\cite{benpas:1998}, but only for the LEP1 energies  
          and for the so-called BARE acceptance.}. 
An important step in this direction was made during the '95 Workshop
``Physics at LEP2''~\cite{lepybk96}, where comparisons of several codes 
for LABH at LEP1 and LEP2 energies were performed. 
They showed, however, that the predictions of various programs can differ
by as much as $2\%$ at LEP1 and $4\%$ at LEP2.
In this section, we  briefly describe our MC event generator for LABH
called BHWIDE and discuss some important cross-checks of the program
as well as comparisons of its predictions with the results of other programs.

BHWIDE is based on the YFS exclusive exponentiation procedure~\cite{yfs:1961},
where all the IR singularities are summed-up to infinite order
and cancelled out properly in the so-called YFS form factor. 
The remaining non-IR residuals, $\bar{\beta}_n^{(l)}$, corresponding to 
the emission of $n$-real photons, are calculated perturbatively up to a given
order $l$, where $l\geq n$, and $(l-n)$ is a number of loops in 
the $\bar{\beta}_n^{(l)}$ calculation. 
In BHWIDE an arbitrary number $n$ of real photons with non-zero $p_T$
are generated according to the YFS MC method of Ref.~\cite{bhl2:1992}.   
The non-IR residuals $\bar{\beta}_n^{(l)}$ are calculated up to 
${\cal O}(\alpha)$, i.e. $\bar{\beta}_0^{(1)}$ and $\bar{\beta}_1^{(1)}$
corresponding to zero-real (one-loop) and one-real (zero-loop) photons,
respectively, are included.
In $\bar{\beta}_0^{(1)}$ we implemented two libraries of the ${\cal O}(\alpha)$
virtual EW corrections: (1) the older one of Ref.~\cite{babamc}, which is
not up to date but can be useful for some tests, and (2) the more recent
one of Ref.~\cite{alibaba}. When the genuine weak corrections are switched off
(or numerically negligible) they are equivalent. 
In $\bar{\beta}_0^{(1)}$ we implemented two independent matrix elements
for single-hard-photon radiation: (1) our calculation~\cite{bhw:1997} 
in terms of helicity amplitudes, and (2) the formula of 
CALKUL~\cite{calkul:1982} for the squared matrix element. 
We have checked that the above two representations agree numerically
up to at least 6 digits on an event-by-event basis.
This constitutes a very important technical cross-check of the implementation
of the hard-photon matrix element in BHWIDE.

The MC algorithm of BHWIDE is based on the algorithm of the program 
BHLUMI for SABH~\cite{bhl2:1992} with a few important modifications:
(1) QED interferences between the electron and positron lines 
(``up-down'' interferences) had to be reintroduced as they are important
in LABH;
(2) the full YFS form factor for the $2\rightarrow 2$ process, including
all $s$-, $t$- and $u$-channels, was implemented~\cite{bhw:1997};
(3) the exact ${\cal O}(\alpha)$ matrix element for the full BHABHA process
was included.
The multiphoton radiation is generated at the low-level MC stage as
for the $t$-channel process, while the $s$-channel as well as all
interferences are reintroduced through appropriate MC weights.
This means that the program is more efficient when the $t$-channel
contribution is dominant, as e.g. at LEP2 energies; however, it proved
to work well also at the $Z$ peak.  

Having all necessary ingredients in the program and an appropriate
algorithm for the MC event generation, we had to subject BHWIDE to several
tests in order to check that it gives correct predictions. 
First, we wanted to know whether it reproduces the small-angle 
limit correctly. 
To this end we compared BHWIDE with BHLUMI~4.04~\cite{bhl4:1996}
whose precision in the SABH regime, as shown in the previous section, 
is under the per-mille level. We found that for the angular acceptance
of $1^{\circ} < \theta_e < 10^{\circ}$ the two programs agree within 
$0.1\%$ (statistical error) 
for both the pure ${\cal O}(\alpha)$ QED corrections and the full
YFS exponentiated cross sections at the energies of $5,\,10,$ and $91.19$~GeV.
Then, we turned to large angles ($40^{\circ} < \theta_e < 140^{\circ}$) and 
compared the pure QED ($Z$-exchange switched off) ${\cal O}(\alpha)$ 
predictions of BHWIDE with the ones of the MC program OLDBIS~\cite{bhl2:1992}
(a modernized version of the program OLDBAB~\cite{oldbab}) whose technical
precision was shown to be at the level of $0.02\%$~\cite{bhl2:1992}.
We found an agreement between these two programs up to $0.05\%$ (stat. error)
for both the BARE and CALO acceptances as defined in Ref.~\cite{lepybk96}
at the $Z$-peak energy. This should also remain valid at LEP2 energies,
since without the $Z$ contribution there is no qualitative difference
in LABH at these two energy regimes.  
The above result is a very important technical cross-check of BHWIDE,
both for the implementation of the ${\cal O}(\alpha)$ QED corrections
and for the correctness of the MC algorithms at large angles 
(the ${\cal O}(\alpha)$ result was extracted in BHWIDE from 
the full multiphoton YFS calculation).

Finally, with all contributions/corrections at ${\cal O}(\alpha)_{exp}^{YFS}$
included, we compared BHWIDE with several MC and semi-analytical programs 
for LABH. These comparisons were first presented in Ref.~\cite{lepybk96},
and then updated in Ref.~\cite{bhw:1997}. 
At LEP1 energies, the program TOPAZ0~\cite{topaz0:1993} is considered
a semi-analytical benchmark for LABH. Its theoretical errors for the
BARE-type acceptance, as recently estimated~\cite{benpas:1998},
are $0.4\%$ ($0.2\%$) at the $Z$ peak and $0.3\%$ ($0.7\%$) on the wings
($\pm 2$~GeV off the peak) for the maximum acollinearity of the final
$e^+$ and $e^-$: $\theta_{acol}^{max} = 10^{\circ}$ ($25^{\circ}$). 
From the tables and figures of 
Refs.~\cite{lepybk96,bhw:1997}, we observe that for the BARE trigger
BHWIDE agrees with TOPAZ0 within $0.4\%$ ($0.55\%$) at the $Z$ peak
and within $0.4\%$ ($0.95\%$) on the wings, for 
$\theta_{acol}^{max} = 10^{\circ}$ ($25^{\circ}$). For the CALO trigger the 
agreement between these two programs is within $0.15\%$ ($0.25\%$)
at the $Z$ peak and $0.2\%$ ($0.55\%$) on the wings.
We can see from the above results that the agreement between the two
calculations improves considerably, as it should, for the CALO-type
acceptance, which is closer to the real experiment.

For LEP2 energies, the comparisons in Refs.~\cite{lepybk96,bhw:1997}
were done for the CALO trigger only. Here, the discrepancies between
various calculations are larger than at LEP1. This can be explained
by the fact that most of the programs were constructed for the $Z$ peak
region, i.e. assuming that the $s$-channel contribution is dominant,
while at the LEP2 energies this is not the case. Here, a program
designed for the $t$-channel process, as e.g. SABSPV~\cite{oreste}, 
should be more reliable. 
The results in Refs.~\cite{lepybk96,bhw:1997} show that BHWIDE is within
$1.5\%$ of SABSPV for the whole LEP2 energy range.
We have also checked that for the BARE trigger BHWIDE agrees with
the semi-analytical code ALIBABA~\cite{alibaba} within $0.3\%$ at the
pure ${\cal O}(\alpha)$ level and within $1\%$ when all corrections are 
included, for the same energy range.

From the above comparisons we can see that BHWIDE is in a better agreement
with the semi-analytical benchmarks for LEP1 and LEP2 energies than
any other MC event generator for LABH.
A more detailed analysis of the BHWIDE theoretical precision is 
in progress now.

\section{Conclusions and Outlook}

We have discussed some aspects of the Bhabha scattering at LEP1 and LEP2.
For the small-angle Bhabha process, we have presented the new error
budget of the program BHLUMI~4.04 based on the exact ${\cal O}(\alpha^2)$
calculations. We have shown that the theoretical error for the luminosity
measurement can be reduced now from $0.11\%$ to $0.061\%$ at LEP1 and
from $0.25\%$ to $0.122\%$ at the LEP2 energy of $176$~GeV. The predictions
of BHLUMI~4.04 remain unchanged. The exact calculations can be
included in the future version of the program if necessary. 
For the large-angle Bhabha process, we have presented the MC event 
generator BHWIDE and discussed some of its cross-checks and comparisons
with other programs. From this we conclude that BHWIDE is the most
precise MC event generator for LABH at LEP1 and LEP2. A comprehensive
analysis of its theoretical errors is in progress.

\section*{Acknowledgements}
Three of the authors (S.J., W.P. and B.F.L.W.) would like to thank 
the CERN TH and EP Divisions and all four LEP Collaborations
for their support.
B.F.L.W. would like to thank Prof. C. Prescott of Group A at SLAC for his 
kind hospitality while this work was in its developmental stages.
This work was supported in part by 
Polish Government grants 
KBN 2P03B08414, 
KBN 2P03B14715, 
the US DoE contracts DE-FG05-91ER40627 and DE-AC03-76SF00515,
and the Maria Sk\l{}odowska-Curie Joint Fund II PAA/DOE-97-316.

\section*{References}


\end{document}